\documentclass[preprint,12pt,a4paper]{elsarticle}
\usepackage{amssymb}
\usepackage{lineno}
\usepackage{float}
\restylefloat{table}
\usepackage[colorlinks=true, urlcolor=blue, pdfborder={0 0 0}]{hyperref}
\usepackage{breakurl}
\usepackage{placeins}
\usepackage{geometry}

\journal{SoftwareX}

\begin{document}
\begin{frontmatter}

\title{\texttt{Davos}: a Python package ``smuggler'' for constructing
  lightweight reproducible notebooks}
\author{Paxton C. Fitzpatrick}
\author{Jeremy R. Manning\corref{cor}}
\ead{Jeremy.R.Manning@Dartmouth.edu}
\cortext[cor]{Corresponding author}
\address{Department of Psychological and Brain Sciences\\Dartmouth College, Hanover, NH 03755}

\begin{abstract}
  Reproducibility is a core requirement of modern scientific research.
  For computational research, reproducibility means that code should
  produce the same results, even when run on different systems. A
  standard approach to ensuring reproducibility entails packaging a
  project's dependencies along with its primary code base. Existing
  solutions vary in how deeply these dependencies are specified,
  ranging from virtual environments, to containers, to virtual
  machines. Each of these existing solutions requires installing or
  setting up a system for running the desired code, increasing the
  complexity and time cost of both sharing and engaging with reproducible
  science. Here, we propose a lighter-weight solution: the
  \texttt{Davos} package. When used in combination with a
  notebook-based Python project, \texttt{Davos} provides a mechanism
  for specifying the correct versions of the project's
  dependencies directly within the code that requires them,
  and automatically installing them in an isolated environment
  when the code is run. The \texttt{Davos} package further
  ensures that these packages and specific versions are used every
  time the notebook's code is executed. This enables researchers to
  share a complete reproducible copy of their code within a single
  Jupyter notebook file.
\end{abstract}


\end{frontmatter}

\section*{Metadata}

\section*{Current code version}

\begin{table}[H]
  \footnotesize
\begin{tabular}{|l|p{6cm}|p{8cm}|}
\hline
\textbf{Nr.} & \textbf{Code metadata description} & \textbf{Metadata value} \\
\hline
C1 & Current code version &  v0.2.3 \\
\hline
C2 & Permanent link to code/repository used for this code version & \url{https://doi.org/10.5281/zenodo.8233890} \\
\hline
C3 & Code Ocean compute capsule & N/A\\
\hline
C4 & Legal Code License & MIT \\
\hline
C5 & Code versioning system used & git \\
\hline
C6 & Software code languages, tools, and services used & Python, JavaScript, PyPI/pip, IPython, Jupyter, ipykernel, PyZMQ.\newline Additional tools used for tests: pytest, Selenium, Requests, mypy, GitHub Actions \\
\hline
C7 & Compilation requirements, operating environments, and
     dependencies & Dependencies:~Python $\geq 3.6$, packaging, setuptools.\newline Supported OSes: MacOS, Linux, Unix-like.\newline Supported IPython environments: Jupyter Notebooks, JupyterLab, Google Colaboratory, Binder, Kaggle, IDE-based notebook editors, IPython shell. \\
\hline
C8 & Link to developer documentation/manual & \url{https://github.com/ContextLab/davos\#readme} \\
\hline
C9 & Support email for questions & \href{mailto:contextualdynamics@gmail.com}{contextualdynamics@gmail.com} \\
\hline
\end{tabular}
\caption{Code metadata}
\label{}
\end{table}


\section{Motivation and significance}

The same computer code may not behave identically under different
circumstances. For example, when code depends on external packages,
different versions of those packages may function differently. Or
when CPU or GPU instruction sets differ across machines, the same
high-level code may be compiled into different machine instructions.
Because executing identical code does not guarantee identical
outcomes, code sharing alone is often insufficient for enabling
researchers to reproduce each other's work, or to collaborate on
projects involving data collection or analysis.

Within the Python~\cite{vanR95} community, external packages that are
published in the most popular repositories~\cite{Pyth03, cond15} are
associated with version numbers and tags that allow users to guarantee
they are installing exactly the same code across different computing
environments~\cite{CoghStuf13}. While it is \textit{possible} to
manually install the intended version of every dependency of a Python
script or package, manually tracking down those dependencies can
impose a substantial burden on the user and create room for mistakes
and inconsistencies. Further, when dependency versions are left
unspecified, replicating the original computing environment becomes
difficult or impossible~\citep{PimeEtal19}.

\begin{figure}[tp]
\centering
\includegraphics[width=0.65\textwidth]{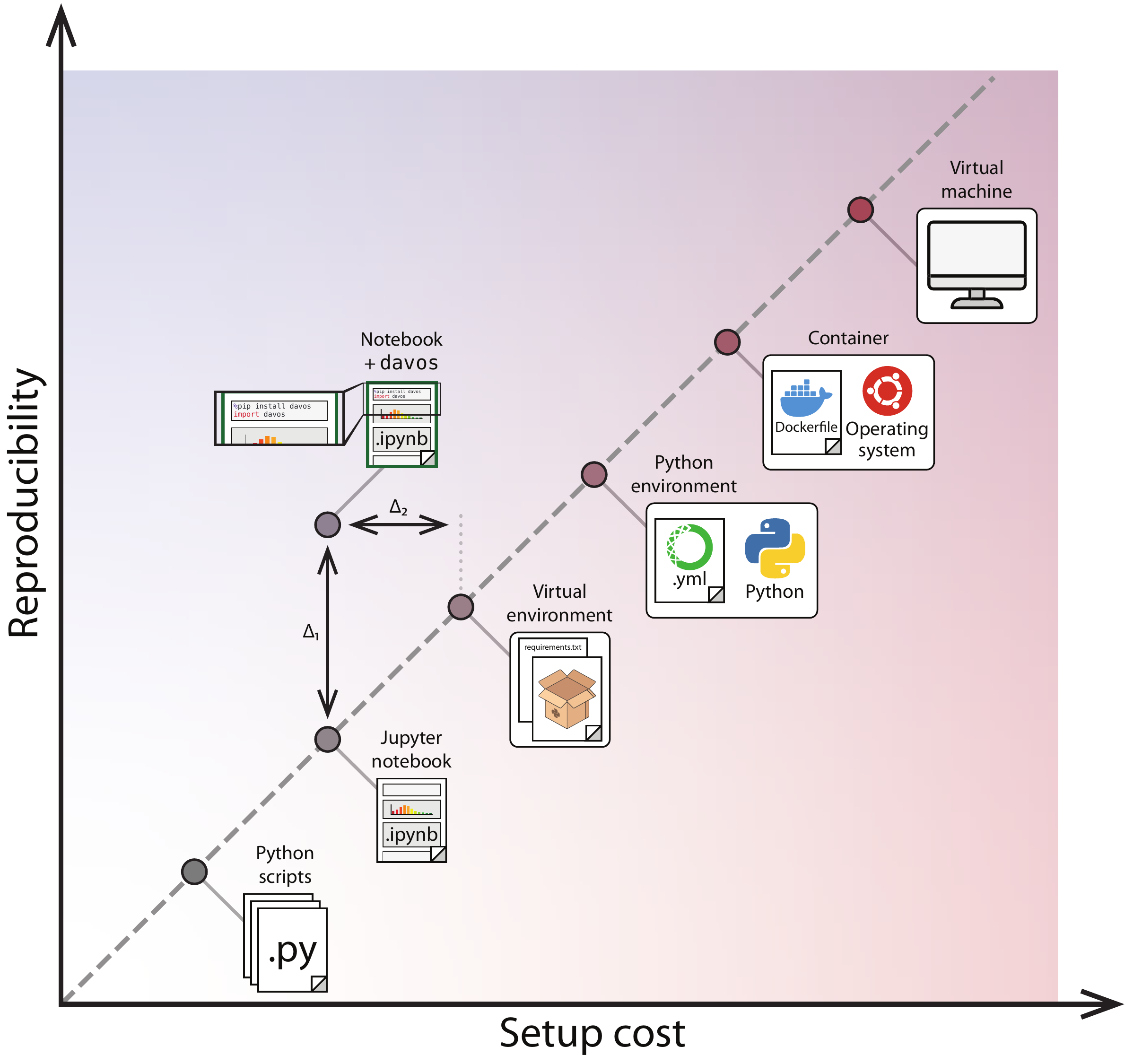}
\caption{\small \textbf{Systems for sharing code within the Python
  ecosystem.} The $x$-axis denotes the ``burden'' placed on users to
  install and configure the given system (systems placed further to the right, that fall within the redder shading, impose a higher setup cost on the user). The $y$-axis denotes the
  degree to which the system guarantees that the code will run
  similarly for different users (systems placed higher up, that fall within the bluer shading, offer stronger guarantees).
  From left to right, bottom to top: plain-text \textbf{Python
  scripts} (\texttt{.py} files) provide the most basic ``system''
  for sharing raw code. Scripts may reference external packages, but
  those packages must be manually installed on other users' systems.
  Further, any checking needed to verify that the correct versions of
  those packages were installed must also be performed manually.
  \textbf{Jupyter notebooks} (\texttt{.ipynb} files) comprise embedded
  text, executable code, and media (including rendered figures, code
  output, etc.). When the \textbf{Davos package} is imported
  into a Jupyter notebook, the notebook's functionality is extended to
  automatically install any required external packages (at their
  correct versions, when specified) into an isolated directory kept separate
  from the user's existing set of packages. \textbf{Virtual environments}
  similarly allow users to install required packages into an isolated
  directory, and \textbf{shareable Python environments} extend this to include
  a specific version of Python itself. However, both of these systems require
  users to create, populate, and manage environments manually. This
  typically entails distributing a configuration file (e.g., a
  \texttt{requirements.txt}, \texttt{pyproject.toml}~\cite{CannEtal16}, or
  \texttt{environment.yml} file) that specifies all project dependencies
  (including version numbers) alongside the primary code base. Users can then
  install a third-party tool~\cite[e.g.,][]{BickEtal07, Eust19, Anac12} to read
  the file and build the environment. \textbf{Containers} provide a means of
  defining an isolated environment that includes a complete operating
  system (independent of the user's operating system), in addition to
  (optionally) specifying a virtual environment or other
  configurations needed to provide the necessary computing
  environment. Containers are typically defined using
  specification files (e.g., a plain-text \texttt{Dockerfile}~\cite{Merk14}) that instruct the
  virtualization engine regarding how to build the containerized
  environment. \textbf{Virtual machines} provide a complete
  hardware-level simulation of the computing environment. In addition
  to simulating specific hardware, virtual machines (typically
  specified using binary image files) must also define operating
  system-level properties of the computing environment. $\Delta_1$
  represents the increased reproducibility \texttt{Davos} provides over standard
  Jupyter notebooks for no greater setup cost. $\Delta_2$ represents
  \texttt{Davos}'s lower setup cost compared to standard virtual environments
  despite more stringently ensuring reproducibility.}
\label{fig:code-sharing}
\end{figure}

Computational researchers and other programmers have de\-vel\-oped a
broad set of approaches and tools to facilitate code sharing and
reproducible outcomes (Fig.~\ref{fig:code-sharing}). At one extreme,
simply distributing a set of Python scripts (\texttt{.py} files) may
enable others to use or gain insights into the relevant work. Because
Python is installed by default on most modern operating systems, for
some projects, this may be sufficient. Another popular approach
entails creating Jupyter notebooks~\cite{KluyEtal16} that comprise a
mix of text, executable code, and embedded media. Notebooks may call
or import external scripts or packages---or even intersperse snippets
of other programming or markup lang\-uages---in order to provide a
more compact and readable experience for users. Both of these systems
(Python scripts and notebooks) provide a convenient means of sharing
code, with the caveat that they do not specify the computing
environment in which the code is executed. Therefore the functionality
of code shared using these systems cannot be guaranteed across
different users or setups.

At another extreme, virtual machines~\cite{Gold74, AltiEtal05, Rose99}
provide a hardware-level simulation of the desired system. Virtual
machines are typically isolated, such that installing or running
software on a virtual machine does not impact the user's primary
operating system or computing environment.
Containers~\cite[e.g.,][]{Merk14, KurtEtal17} provide a similar
``isolated'' experience. Although containerized environments do not
specify hardware-level operations, they are typically packaged with a
complete operating system, in addition to a complete copy of Python
and any relevant package dependencies. Shareable Python
environments~\cite[e.g.,][]{Anac12} also provide a computing
environment that is largely separated from the user's main
environment. They incorporate a copy of Python and the target
software's dependencies, but do not specify or
reproduce an operating system for the runtime environment.
Virtual environments~\cite[e.g.,][]{BickEtal07, Eust19} work similarly, but
reuse an existing copy of Python rather than bundling their own.
Each of these systems (virtual machines, containers, Python environments, and virtual environments)
guarantees (to differing degrees---at the hardware level, operating
system level, Python environment level, and package environment level, respectively) that the
relevant code will run similarly for different users. However, each of
these systems also relies on additional software that can be complex
or resource-intensive to install and use, creating potential barriers
to both contributing to and taking advantage of open science
resources.

We designed \texttt{Davos} to occupy a ``sweet spot'' within this space.
\texttt{Davos} is a notebook-installable package that adds functionality to the
default notebook experience. Like standard Jupyter notebooks,
\texttt{Davos}-enhanced notebooks allow researchers to include text, executable
code, and media within a single file. No further setup or installation is
required from the user, beyond what is needed to run a standard Jupyter
notebook. And like virtual environments, \texttt{Davos} provides a convenient
mechanism for fully specifying (and installing, as needed) a complete set of
Python dependencies, including specific package versions, which are contained
and isolated from the rest of the user's system.

\section{Software description}

The \texttt{Davos} package is named after Davos Seaworth, a smuggler referred
to as ``the Onion Knight'' from the series \textit{A Song of Ice and Fire} by
George R. R. Martin~\cite{Mart98}. The \texttt{smuggle} keyword provided by
\texttt{Davos} is a play on Python's \texttt{import} keyword: whereas importing
can load a package into the Python workspace within the existing rules and
frameworks provided by the Python language, ``smuggling'' provides an
alternative that expands the scope and reach of ``importing.'' Like the
character Davos Seaworth (who became famous for smuggling onions through a
blockade on his homeland), the \texttt{Davos} package uses ``onion comments'' to precisely control how
packages are smuggled into the Python workspace.

\begin{figure}[tp]
\centering
\includegraphics[width=\textwidth]{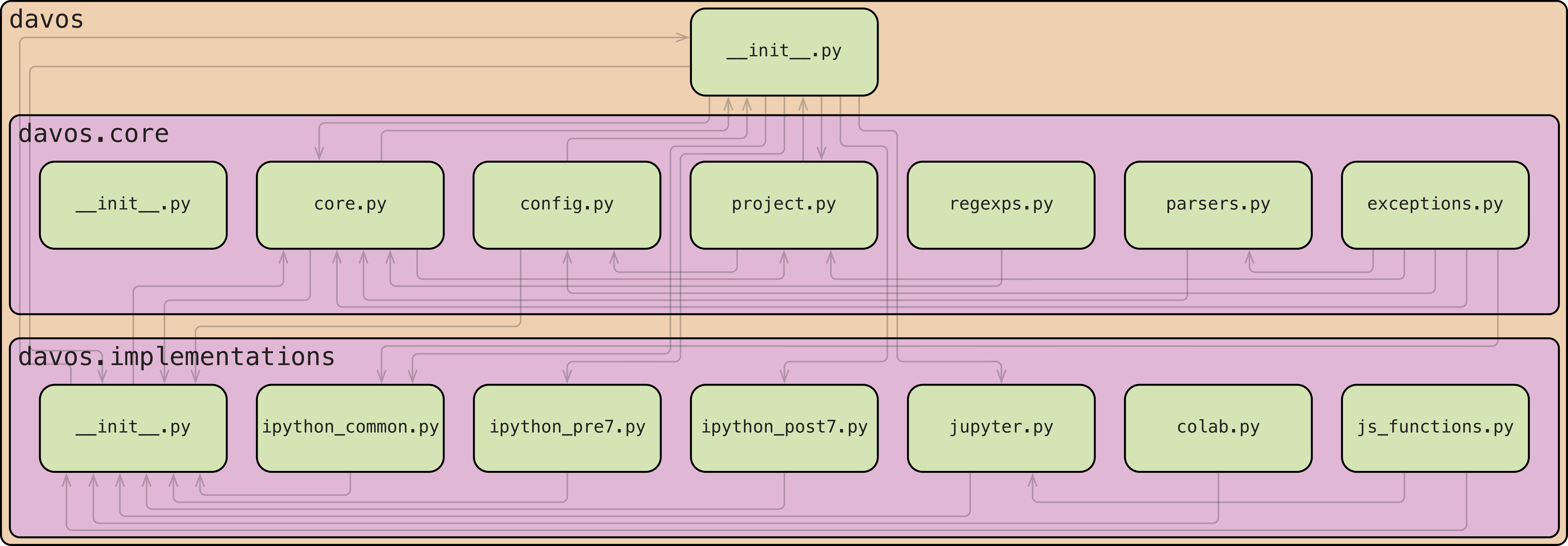}
\caption{\small \textbf{Package structure.} The \texttt{Davos} package
  comprises two interdependent subpackages. The \texttt{davos.core}
  subpackage includes modules for parsing \texttt{smuggle} statements
  and onion comments, installing and validating packages, isolating and managing
  installed packages, and configuring \texttt{Davos}'s behavior. The
  \texttt{davos.implementations} subpackage includes
  environment-specific modifications and features that are needed to
  support the core functionality across different notebook-based
  environments. Individual modules (i.e., \texttt{.py} files) are represented by lime
  rounded rectangles, and arrows denote dependencies (each arrow
  points to a module that imports objects defined in the module at the
  arrow's source).}
\label{fig:package-structure}
\end{figure}

\subsection{Software architecture}\label{sec:architecture}

The \texttt{Davos} package consists of two interdependent subpackages
(see Fig.~\ref{fig:package-structure}). The first,
\texttt{davos.core}, comprises a set of modules that
implement the bulk of the package's core
functionality, including pipelines for installing and validating
packages, custom parsers for the \texttt{smuggle} statement (see
Sec.~\ref{subsec:smuggle}) and onion comment (see
Sec.~\ref{subsec:onion}), a system for isolating dependencies of
different projects (see Sec.~\ref{subsec:projects}), and a runtime
interface for configuring \texttt{Davos}'s behavior (see Sec.~\ref{subsec:config}).
However, certain critical aspects of this
functionality require (often substantially) different implementations
depending on properties of the notebook environment in which
\texttt{Davos} is used (e.g., whether the frontend is provided by
Jupyter or Google Colaboratory, or which version of
IPython~\cite{PereGran07} is used by the notebook kernel). To deal
with this, environment-dependent components of core features and behaviors
are isolated and abstracted to ``helper functions'' in the
\texttt{davos.implementations} subpackage. This second subpackage
defines multiple, interchangeable versions of each helper function,
organized into modules by the conditions that trigger their use. At
runtime, \texttt{Davos} detects various features in the notebook
environment and selectively imports a single version of each helper
function into the top-level \texttt{davos.implementations} namespace,
allowing \texttt{davos.core} modules to access the proper
implementations for the current notebook environment in a single,
consistent location. An additional benefit of this design is that it
allows maintainers and users to extend \texttt{Davos} to
support new, updated, or custom notebook variants by adding new
\texttt{davos.implementations} modules that define their own versions
of each helper function, modified from existing implementations as
needed.

\subsection{Software functionalities}

\subsubsection{The \texttt{smuggle} statement}\label{subsec:smuggle}

Functionally, importing \texttt{Davos} in an IPython notebook enables
an additional Python keyword: ``\texttt{smuggle}'' (see
Sec.~\ref{subsec:implementation} for details on how this works).
The \texttt{smuggle} keyword can be used as a drop-in
replacement for Python's built-in \texttt{import} keyword to load
packages, modules, and other objects into the notebook's namespace.
However, whereas \texttt{import} will fail if the requested package is
not installed locally, \texttt{smuggle} statements can handle missing
packages on the fly. If a smuggled package does not exist in the
user's Python environment, \texttt{Davos} will download and install it automatically,
expose its contents to Python's \texttt{import} machinery, and load it
into the notebook for immediate use.

Importantly, packages installed by \texttt{Davos} are made available for use in the
notebook without affecting the user's Python environment or existing packages.
By default, \texttt{smuggle} statements will install missing packages (and any
missing dependencies of those packages) into a notebook-specific, virtual
environment-like directory called a ``project'' (see
Sec.~\ref{subsec:projects}). In turn, \texttt{smuggle} statements executed in a
particular notebook will preferentially load packages from that notebook's
project directory whenever they are available, rather than searching for them
in the user's main Python environment. In this way, \texttt{smuggle}
statements can be substituted for \texttt{import} statements to automatically
ensure that all packages needed to run a notebook are installed and available
at runtime each time the notebook is run, without risking interfering with
dependencies of the user's other Python programs, or other \texttt{Davos}-enhanced
notebooks.

\subsubsection{The onion comment}\label{subsec:onion}

For greater control over the behavior of \texttt{smuggle} statements, \texttt{Davos}
defines an additional construct called the ``onion comment.'' An onion comment
is a special type of inline comment that may be placed on a line containing a
\texttt{smuggle} statement to customize how \texttt{Davos} searches for the smuggled
package locally and, if necessary, downloads and installs it. Onion comments
follow a simple format based on the ``type comment'' syntax introduced in PEP
484~\cite{vanREtal14}, and are designed to make managing packages with \texttt{Davos}
intuitive and familiar. To construct an onion comment, users provide the name
of the installer program (e.g., \texttt{pip}) and the same arguments one would
use to manually install the package as desired via the command line:
\begin{center}
  \includegraphics[width=0.9\textwidth]{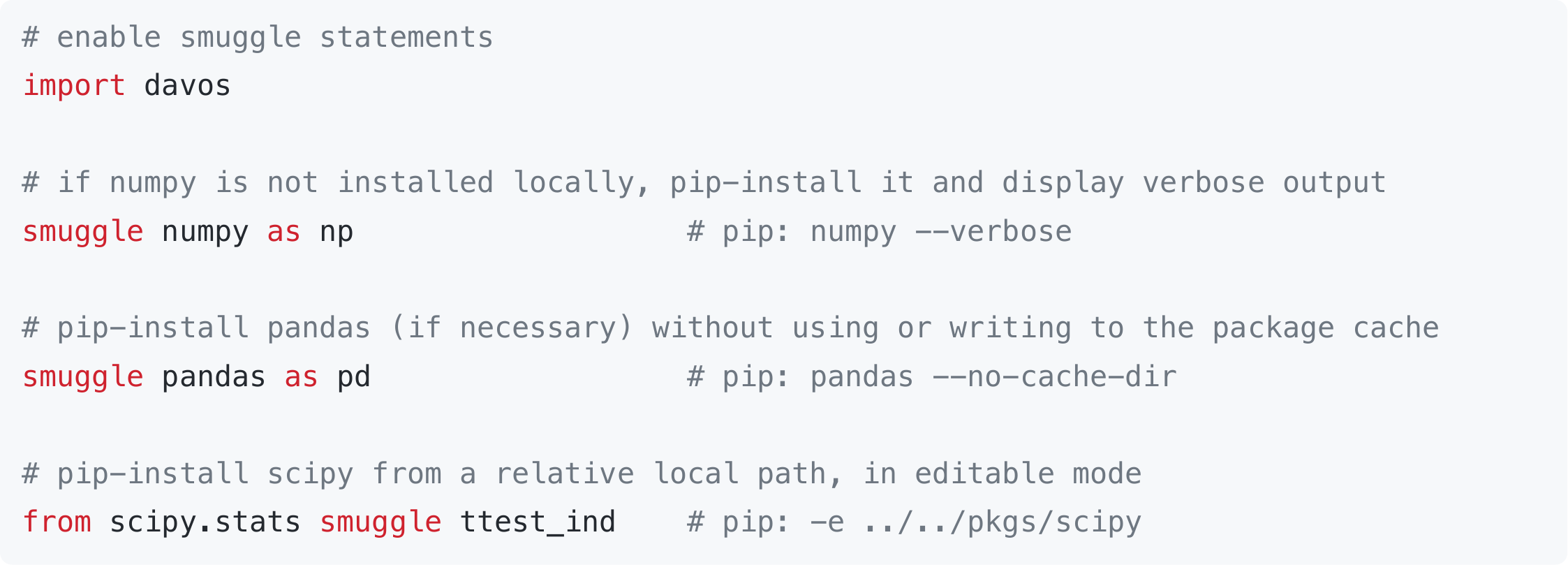}
\end{center}
Occasionally, a package's distribution name (i.e., the name used
when installing it) may differ from its top-level module name (i.e., the name
used when importing it). In such cases, an onion comment can be used to ensure
that \texttt{Davos} installs the proper package if it cannot be found locally:
\begin{center}
  \includegraphics[width=0.9\textwidth]{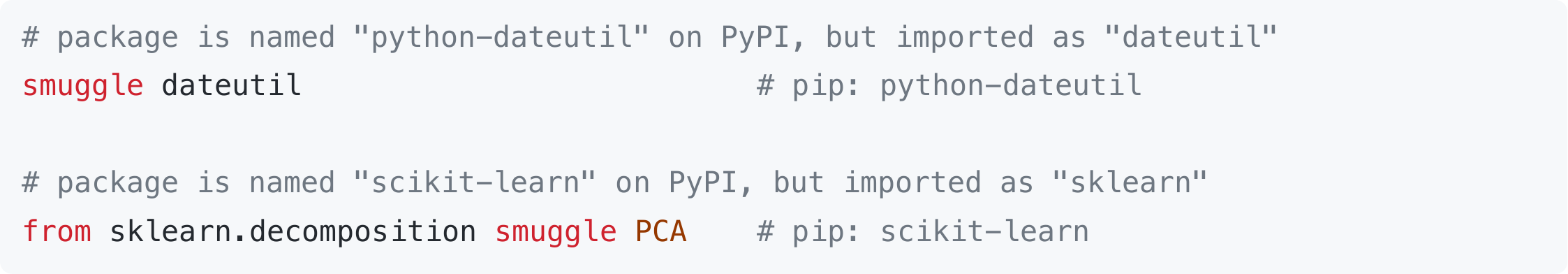}
\end{center}
Because onion comments may be constructed to specify any aspect of
the installer program's behavior, they provide a mechanism for precisely controlling
how, where, and when smuggled packages are installed. Critically, if an onion
comment includes a version specifier~\cite{CoghStuf13}, \texttt{Davos} will ensure that
the version of the package loaded into the notebook matches the specific
version requested (or satisfies the given version constraints). If the smuggled
package exists locally, \texttt{Davos} will extract its version information from its
metadata and compare it to the specifier provided. If the two are incompatible
(or no local installation is found), \texttt{Davos} will download, install, and load a
suitable version of the package instead:
\begin{center}
  \includegraphics[width=0.9\textwidth]{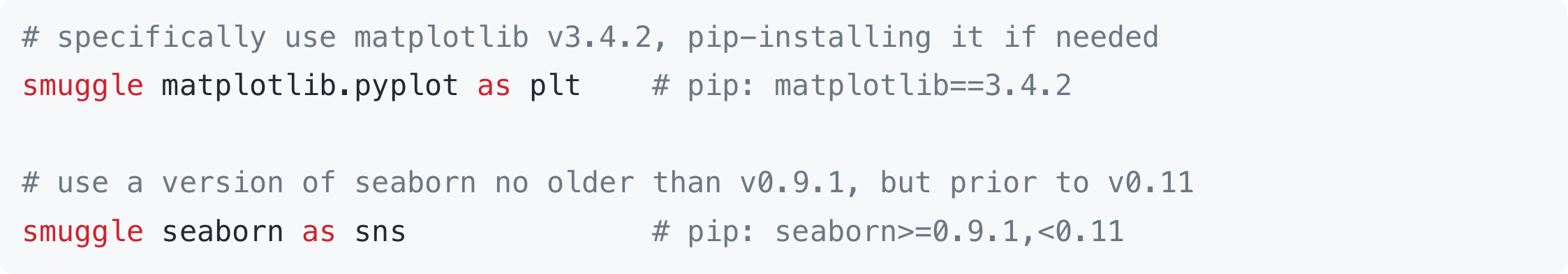}
\end{center}
Onion comments can also be used to \texttt{smuggle} specific VCS references (e.g.,
Git~\cite{TorvHama05} branches, commits, tags, etc.):
\begin{center}
  \includegraphics[width=0.9\textwidth]{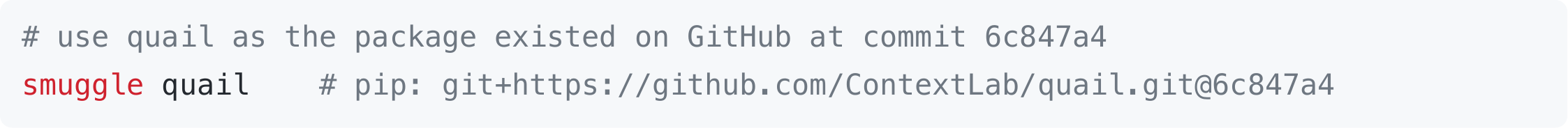}
\end{center}
\texttt{Davos} processes onion comments internally before forwarding arguments to the
installer program. In addition to preventing shared notebooks from executing
arbitrary code in a user's shell, this enables \texttt{Davos} to adjust its behavior
based on how particular flags will affect the behavior of the installer
program. For example, including \texttt{pip}'s \texttt{--no-input} flag will also
temporarily enable \texttt{Davos}'s non-interactive mode (see Sec.~\ref{subsec:config}).
Similarly, if an onion comment contains either \texttt{-I}/\texttt{--ignore-installed},
\texttt{-U}/\texttt{--upgrade}, or \texttt{--force-reinstall}, \texttt{Davos} will
install and load a new copy of the smuggled package without first checking
for it locally:
\begin{center}
  \includegraphics[width=0.9\textwidth]{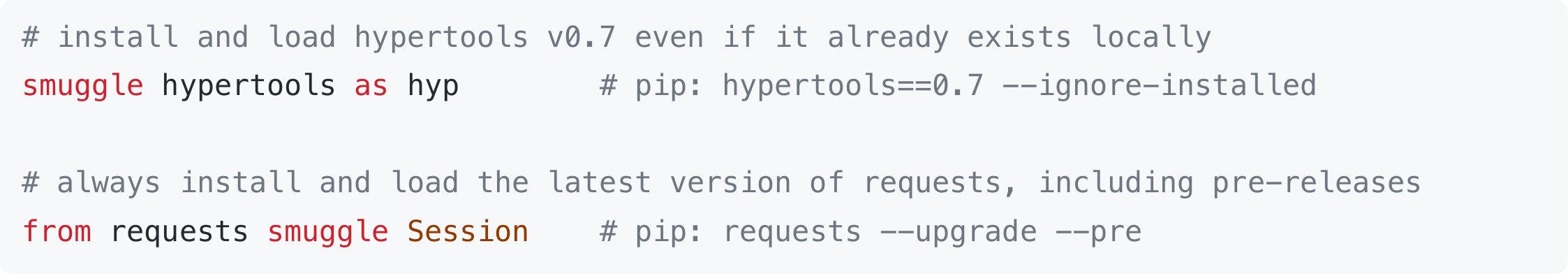}
\end{center}
Since the purpose of an onion comment is to describe how a smuggled package
should be installed (if necessary) so that it can be loaded and used
immediately, options that would normally cause the package not to be installed
(such as \texttt{-h}/\texttt{--help} or \texttt{--dry-run}) are disallowed. Additionally,
when using a \texttt{Davos} ``project'' to isolate smuggled packages (the default behavior;
see Sec.~\ref{subsec:projects}), onion comments may not contain options that
would change the package's installation location (such as
\texttt{-t}/\texttt{--target}, \texttt{--root}, or \texttt{--prefix}). However, if
the user disables project-based isolation and specifies \texttt{--target <dir>},
\texttt{Davos} will ensure that \texttt{<dir>} is included in the module search path (i.e.,
\texttt{sys.path}), prepending it if necessary, so the package can be loaded.

\subsubsection{Projects}\label{subsec:projects}

Standard approaches to installing packages from within a notebook can alter the local Python environment in potentially unexpected and undesired ways.
For example, running a notebook that installs its dependencies via system shell commands (prefixed with ``\texttt{!}'') or IPython magic commands (prefixed with ``\texttt{\%}'') may cause other existing packages in the user's environment to be replaced with alternate versions.
This can lead to incompatibilities between installed packages, affect the behavior of the user's other scripts or notebooks, or even interfere with system applications.

To prevent \texttt{Davos}-enhanced notebooks from having unwanted side effects on the user's environment, any packages installed via \texttt{smuggle} statements are automatically isolated using a custom, virtual environment-like system called ``projects.''
\texttt{Davos} projects are similar to standard Python virtual environments (e.g., created with the standard library's \texttt{venv} module or a third-party tool like \texttt{virtualenv}~\cite{BickEtal07}) but with a few noteworthy differences that make them generally lighter-weight and simpler to use.
Like a standard virtual environment, a \texttt{Davos} project consists of a directory (within a hidden \texttt{.davos} folder in the user's home directory) that houses third-party packages needed for a particular Python project, workflow, or task.
However, unlike standard virtual environments, \texttt{Davos} projects do not need to be manually created, activated, or deactivated, and they function to \textit{extend} the user's existing Python environment rather than replace it.

When \texttt{Davos} is imported into a notebook, a project directory for that notebook is automatically created (if it does not exist already).
When \texttt{smuggle} statements within that notebook are executed, any packages (or specific versions of packages) that are not already available in the user's Python environment are installed into the notebook's project directory (along with any missing dependencies of those packages).
During each \texttt{smuggle} statement's execution, \texttt{Davos} also temporarily prepends the notebook's project directory to the module search path so that these project-installed packages are visible when searching for smuggled packages locally, and prioritized over those in the user's main environment.

Thus, rather than constructing fully separate Python environments from scratch, \texttt{Davos} projects work by supplementing the user's runtime environment with any additional packages (or specific package versions) needed to satisfy the dependencies of their corresponding notebooks.
In some cases, this might include every package smuggled into a notebook (e.g., if the notebook is run inside a freshly created, empty virtual environment).
In other cases, the user's environment may already provide all required packages, and the notebook's project directory will go unused (in which case it will be deleted automatically when the notebook kernel is shut down).
Regardless of the extent to which the existing environment is augmented, \texttt{Davos}'s project system ensures that all smuggled packages are installed locally and loaded successfully at runtime, while the contents of the user's Python environment are never altered.

Because \texttt{smuggle} statements in a given notebook are evaluated every time the notebook is run, this ensures that the notebook's requirements will remain satisfied even if the user's Python environment changes.
For example, suppose a user has \texttt{NumPy}~\cite{HarrEtal20} v1.24.3 installed in their current Python environment and runs a \texttt{Davos}-enhanced notebook that smuggles \texttt{NumPy} with ``\texttt{numpy==1.24.3}'' specified in an onion comment (see Sec.~\ref{subsec:onion}).
Since the user's existing version of the package satisfies this requirement, \texttt{Davos} will load it into the notebook's runtime environment.
But if the user later upgrades their environment's \texttt{NumPy} version to v1.25.0 (perhaps as a result of installing a different package that depends on it) and subsequently re-runs this notebook, the local version will longer satisfy this requirement, so \texttt{Davos} will install \texttt{NumPy} v1.24.3 into the notebook's project directory and load that version instead.
From then on, any further changes to the user's \texttt{Numpy} installation would have no effect on \texttt{Davos}'s behavior in this particular notebook, as a satisfactory version now exists in its project directory.
(If the version specified in the onion comment were changed, \texttt{Davos} would update the version installed in the project directory accordingly.)
For efficiency, \texttt{Davos} projects will generally not duplicate dependencies already satisfied by the user's Python environment.
However, if desired, adding \texttt{pip}'s \texttt{--ignore-installed} flag to an onion comment in the notebook will cause \texttt{Davos} to install the smuggled package into the project directory whether or not it already exists locally.

By default, each \texttt{Davos}-enhanced notebook will create and use its own notebook-specific project named for the absolute path to the notebook file.
However, before smuggling its required packages, a notebook may be set to instead use an arbitrarily named, notebook-agnostic project by assigning any (non-empty) string to \texttt{davos.project} (see Sec.~\ref{subsec:config}).
This provides a convenient way for multiple related notebooks that share a common set of requirements to use the same \texttt{Davos} project, by setting \texttt{davos.project} to the same string in each one.
It is also possible (though typically not recommended) to disable \texttt{Davos}'s project system and instead install smuggled packages directly into the user's Python environment by setting \texttt{davos.project} to \texttt{None}.

When accessed (unless its value has been set to \texttt{None}), \texttt{davos.project} will evaluate to a \texttt{Project} object that represents the project used by the current notebook (strings assigned to \texttt{davos.project} are converted to \texttt{Project}s internally). This object supports methods for interacting with the current project, including locating its directory within the file system, listing all installed packages' names and versions, changing the project's name, and deleting its contents.
\texttt{Project} instances can also be created and managed programmatically, and \texttt{Davos} provides additional utilities for viewing and working with all existing projects (see Secs.~\ref{subsec:config} and~\ref{subsec:toplevel}).

\subsubsection{Configuring and querying \texttt{Davos}}\label{subsec:config}

After importing \texttt{Davos} into a notebook, the top-level \texttt{davos} module exposes a set of attributes whose values determine various aspects of \texttt{Davos}'s behavior.
The majority of these are writeable options that can be modified to customize how, where, and when \texttt{Davos} installs smuggled packages (see Sec.~\ref{sec:illustrative-example} for an illustrative example).
These include:

\begin{itemize}
\item \texttt{.active}: This attribute controls whether support for \texttt{smuggle}
  statements and onion comments is enabled (\texttt{True}) or
  disabled (\texttt{False}). When \texttt{Davos} is first imported,
  \texttt{davos.active} is set to \texttt{True} (see Sec.~\ref{subsec:implementation} for implementation details and additional information).

\item \texttt{.auto\_rerun}: This attribute controls how
  \texttt{Davos} behaves when attempting to \texttt{smuggle} a new
  version of a package that was previously loaded (via an \texttt{import} or \texttt{smuggle} statement) and cannot be
  reloaded. This can happen if the package includes extension modules
  that dynamically link C or C++ objects to the Python interpreter,
  and the code that generates those objects was changed between the
  previously loaded and to-be-smuggled versions. If this attribute
  is set to \texttt{True}, \texttt{Davos} will automatically restart
  the notebook kernel and re-run all code up to (and including) the
  current \texttt{smuggle} statement. If set to \texttt{False} (the default),
  \texttt{Davos} will instead issue a warning, pause execution, and
  prompt the user to either restart and re-run the notebook, or
  continue running with the previously loaded package version until
  the next time the kernel is restarted manually. Note that, as of
  this writing, setting \texttt{davos.auto\_rerun} to \texttt{True} is not
  supported in Google Colaboratory notebooks.

\item \texttt{.confirm\_install}: If set to \texttt{True} (default:
  \texttt{False}), \texttt{Davos} will require user confirmation
  before installing a smuggled package that is not already
  available locally. This is primarily useful if the user has disabled
  \texttt{Davos}'s ``project'' system for isolating smuggled packages (see
  Sec.~\ref{subsec:projects}) but still wants to carefully control what
  packages are installed into their main Python environment.

\item \texttt{.noninteractive}: Setting this attribute to
  \texttt{True} (default: \texttt{False}) enables non-in\-ter\-act\-ive
  mode, in which all user interactions (prompts and dialogues) are
  disabled. Note that in non-interactive mode, the
  \texttt{confirm\_install} option is set to \texttt{False}. If
  \texttt{auto\_rerun} is set to \texttt{False} while in non-interactive
  mode, \texttt{Davos} will raise an exception if a smuggled package
  cannot be reloaded, rather than prompting the user.

\item \texttt{.pip\_executable}: This attribute's value specifies the
  path to the \texttt{pip} executable used to install smuggled
  packages. The default is programmatically determined from the user's Python
  environment and falls back to \texttt{<sys.executable> -m pip} if no
  executable can be found.

\item \texttt{.project}: This attribute's value is a \texttt{Project} instance representing the \texttt{Davos} project associated with the current notebook.
  As described in Section~\ref{subsec:projects}, \texttt{Davos} projects serve to isolate packages installed by \texttt{smuggle} statements from the user's main Python environment, and the \texttt{Project} class provides an interface for inspecting and managing projects at runtime.
  This attribute's default value is a notebook-specific project named for the absolute path to the notebook file.
  To change the project used in the current notebook (e.g., in order to use the same project in multiple related notebooks), this attribute may be assigned a different \texttt{Project} instance or, for simplicity, the name of the desired project as a string or \texttt{pathlib.Path} (either of which will be converted to a \texttt{Project} on assignment).
  Alternatively, setting \texttt{davos.project} to \texttt{None} will disable project-based isolation for the current notebook and cause \texttt{Davos} to install any missing packages directly into the main Python environment.
  This attribute can be reset to its default value using the top-level \texttt{use\_default\_project()} function (see Sec.~\ref{subsec:toplevel}).
  For more information about \texttt{Davos} projects, see Section~\ref{subsec:projects}.

\item \texttt{.suppress\_stdout}: If this attribute is set to
  \texttt{True} (default: \texttt{False}), \texttt{Davos} suppresses
  printed (console) outputs from both itself and the installer program.
  This can be useful when smuggling packages that require installing many
  dependencies and/or generate extensive output when built from source
  distributions. Note that if this option is enabled and the installer
  program throws an error, both its stdout and stderr streams will still be
  displayed alongside the Python traceback to allow for debugging.

\end{itemize}

\noindent The attributes above can be modified directly or via the \texttt{davos.configure()} function, which allows setting multiple options simultaneously (see Sec.~\ref{subsec:toplevel} for more information or Sec.~\ref{sec:illustrative-example} for example usage).
In addition to these writeable options, the top-level \texttt{davos} module also provides several read-only attributes that can be displayed in the notebook or checked programmatically at runtime, and contain potentially useful information about the notebook environment or \texttt{Davos}'s internal state:

\begin{itemize}

\item \texttt{.all\_projects}: This attribute contains a list of all \texttt{Davos} projects that exist on the user's local system (see Sec.~\ref{subsec:projects} for more information about \texttt{Davos} projects).
  Each item in this list is either a \texttt{Project} or \texttt{AbstractProject} instance.
  \texttt{Abstract\-Project}s represent notebook-specific projects whose associated notebooks no longer exist.
  They support the same functionality as \texttt{Project} objects (including methods for inspecting, renaming, and deleting them) and serve primarily to help users identify and clean up extraneous projects left behind after deleting \texttt{Davos}-enhanced notebooks (e.g., see Sec.~\ref{subsec:toplevel}).

\item \texttt{.environment}: This attribute's value is a string denoting the set of environment-dependent ``helper functions'' used by \texttt{Davos} in the current notebook.
  As described in Section~\ref{sec:architecture}, \texttt{Davos} internally chooses between interchangeable implementations of certain core features based on various properties of the notebook's frontend and IPython kernel.
  As of this writing, three unique combinations of helper functions are required to support existing notebook environments, ergo this attribute has three possible values: \texttt{"IPython<7.0"}, \texttt{"IPython>=7.0"}, or \texttt{"Colaboratory"}.
  However, this attribute could take on additional values in the future as new notebook interfaces are created and IPython's internals are updated, and as additional versions of helper functions are added to \texttt{Davos} to support them.

\item \texttt{.ipython\_shell}: This attribute contains the global IPython \texttt{InteractiveShell} instance underlying the notebook kernel session.

\item \texttt{.smuggled}: This attribute's value is a Python dictionary that functions as a cache of \texttt{smuggle} statements executed during the current notebook kernel session.
  The dictionary's keys are names of smuggled packages, and its values are arguments passed to the installer program via onion comments.
  Entries appear in order of the \texttt{smuggle} statements' execution.

\end{itemize}

\noindent The current values of all \texttt{davos} attributes may be viewed at once within a notebook by printing the \texttt{davos.config} object.

\subsubsection{Other top-level \texttt{Davos} functions}\label{subsec:toplevel}

The \texttt{Davos} package also provides a handful of functions available in the top-level \texttt{davos} namespace.
Some of these functions serve primarily as conveniences, while others provide additional functionality:

\begin{itemize}

\item \texttt{configure(**kwargs)}: This function provides an alternate way of assigning values to the writeable attributes listed in Section~\ref{subsec:config} and can be used to configure multiple options at once (see Sec.~\ref{sec:illustrative-example} for example usage).
  The function accepts attribute names as keyword-only arguments to which their desired values are passed.
  If any of the options passed are incompatible (e.g., both \texttt{confirm\_install=True} and \texttt{noninteractive=True} are passed) or assignment to any of the specified attributes fails for any reason, none of the given options will be modified.

\item \texttt{get\_project(project\_name, create=False)}: This function can be passed the name of a \texttt{Davos} project (\texttt{project\_name}) to get the \texttt{Project} or \texttt{AbstractProject} instance representing it.
  The optional \texttt{create} argument determines the function's behavior when no project with the given name exists: if \texttt{create=False} (the default), the function will return \texttt{None}; if \texttt{create=True}, a project with the given name will be created and returned.

\item \texttt{prune\_projects(yes=False)}: This function allows users to quickly ``clean up'' their local \texttt{Davos} projects by deleting notebook-specific projects whose corresponding notebooks no longer exist (i.e., \texttt{AbstractProject}s).
  As with standard virtual environments, periodically removing unused project directories can be useful for reclaiming disk space from dependencies of code that is no longer in use.
  By default, this function will interactively display a list of all unused projects and allow the user to choose whether or not to delete each one.
  Alternatively, passing \texttt{yes=True} will immediately remove all unused projects without prompting for confirmation.
  Note that if \texttt{Davos}'s non-interactive mode is enabled (see Sec.~\ref{subsec:config}), \texttt{yes=True} must be explicitly passed, otherwise the function will raise an exception.
  This serves as a safeguard against accidentally deleting projects, since non-interactive mode disables all user input and confirmation.
  Also note that this function will not delete notebook-agnostic projects (i.e., manually created projects whose names are not notebook file paths), as they are not linked to specific notebooks whose existence determines whether or not they are still needed.
  These (and any) projects may be deleted individually by calling their \texttt{Project} objects' \texttt{.remove()} method.

\item \texttt{require\_python(version\_spec, warn=False, extra\_msg=None, pre\-re\-leases=\\None)}: Through \texttt{smuggle} statements and onion comments, \texttt{Davos} can automatically ensure that all Python packages needed to run a notebook are installed, and that the same versions of those packages are used no matter when or by whom the notebook is run.
  However, because \texttt{Davos} operates at runtime, one thing it cannot do automatically is install and switch to a specific version of Python itself.
  Distributing shared code along with a precise Python version for running it requires a heavier-weight solution, such as a Conda environment or Docker container (see Fig.~\ref{fig:code-sharing}).
  Yet a \texttt{Davos}-enhanced notebook may still \texttt{smuggle} certain packages that depend on users having a particular Python version or range of versions (e.g., even just within the standard library, the \texttt{dataclass} module was first added in Python 3.7~\cite{Smit17} and at least 19 modules are slated for removal in Python 3.13~\cite{HeimCann19}).
  The \texttt{davos.require\_python()} function can be added to the top of a \texttt{Davos}-enhanced notebook to communicate to users that the notebook's code should be run with a specific or constrained Python version (see Sec.~\ref{sec:illustrative-example} for example usage).
  The function may be passed a version identifier (e.g., \texttt{"3.10.5"}) or any valid version specifier~\cite{CoghStuf13} (e.g., \texttt{"\raisebox{0.5ex}{\texttildelow}=3.11"}, \texttt{">=3.9;<3.12"}, etc.) and will raise an exception if the user's Python version is incompatible.
  Alternatively, a ``soft'' or suggested constraint can be imposed by passing \texttt{warn=True} to issue a warning rather than raise an error.
  Additional information can be added to the default error/warning message (e.g., the specific reason for this requirement) via the \texttt{extra\_msg} argument, and the optional \texttt{prereleases} argument can be used to explicitly allow (\texttt{True}) or disallow (\texttt{False}) pre-release versions (by default, the policy is determined by the value of \texttt{version\_spec}).

\item \texttt{use\_default\_project()}: By default, each \texttt{Davos}-enhanced notebook will create and use a notebook-specific project named based on its absolute path.
  If a user manually changes the project used by the current notebook (i.e., by setting the value of the \texttt{davos.project} attribute; see Sec.~\ref{subsec:config}), this function can be called to switch back to using the notebook's default project and reset \texttt{davos.project} to its default value.
  See Section~\ref{subsec:projects} for more information about \texttt{Davos} projects and Section~\ref{sec:illustrative-example} for an illustrative example.

\end{itemize}

\subsection{Implementation details}\label{subsec:implementation}

Although \texttt{Davos} is designed to \textit{appear} to add a new
keyword to Python's vocabulary, this illusion is actually created through
several ``hacks'' that make use of the notebook's IPython backend
for processing and executing users' code. Specifically, when
\texttt{Davos} is first imported, or when it is activated after having been
set to an inactive state, two actions are triggered. First, the
\texttt{smuggle()} function is injected into the IPython user
namespace. Second, the \texttt{Davos} parser is registered as a
custom IPython input transformer.

\begin{figure}[tp]
\centering
\includegraphics[width=\textwidth]{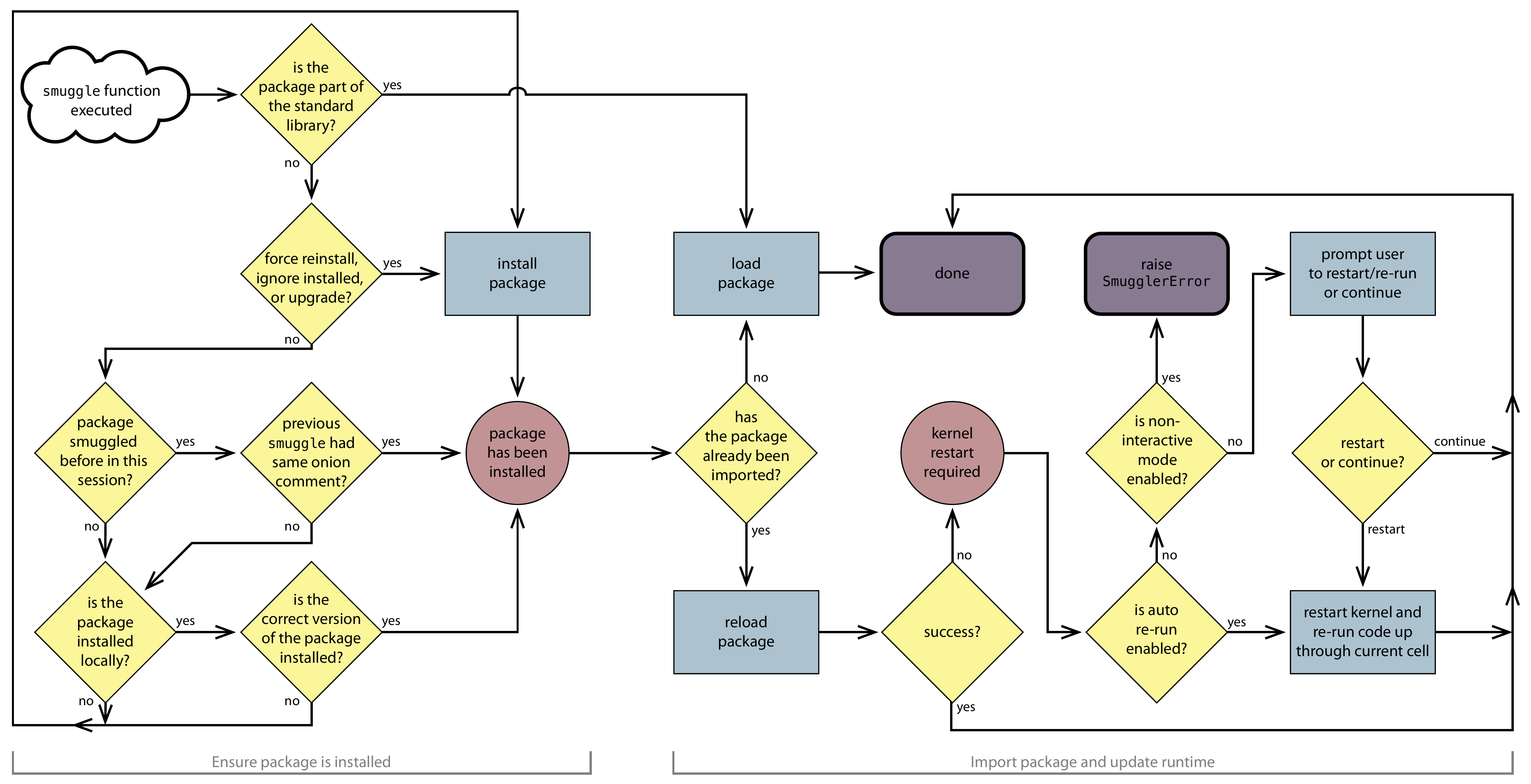}
\caption{\small \textbf{\texttt{smuggle()} function algorithm.} At a
  high level, the \texttt{smuggle()} function may be conceptualized as
  following two basic steps. First (left), \texttt{Davos} ensures that the
  correct version of the desired package is available locally, installing
  it automatically (into the notebook's project directory) if needed. Second (right),
  \texttt{Davos} loads the package into the notebook and updates the current
  runtime environment.}
\label{fig:flow-chart}
\end{figure}

IPython preprocesses all executed code as plain text before it is sent
to the Python compiler, in order to handle special constructs like
\mbox{\texttt{!}-pre}\-fixed shell commands and \mbox{\texttt{\%}-pre}\-fixed ``magic'' commands. \texttt{Davos} uses
this same process to invisibly transform \texttt{smuggle} statements into
syntactically valid Python code. The \texttt{Davos} parser uses a
regular expression to match lines of code containing \texttt{smuggle}
statements (and, optionally, onion comments), extract relevant
information from their text, and replace them with equivalent calls to
the \texttt{smuggle()} function. For example, if a user runs a
notebook cell containing
\begin{center}
\includegraphics[width=0.9\textwidth]{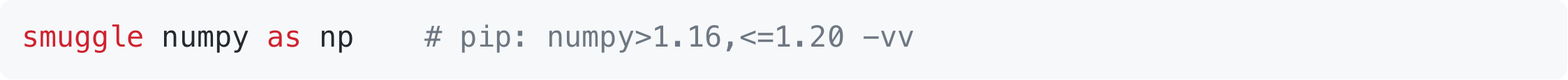}
\end{center}
the code that is actually executed by the Python interpreter would be
\begin{center}
\includegraphics[width=0.9\textwidth]{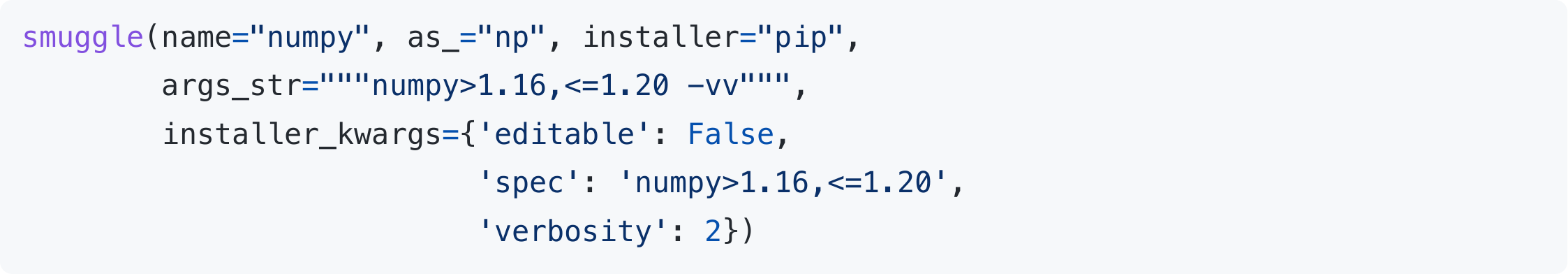}
\end{center}
The call to the \texttt{smuggle()} function carries out
\texttt{Davos}'s central logic by determining whether the smuggled
package must be installed, carrying out the installation if necessary,
and subsequently loading it into the namespace. This process is
outlined in Figure~\ref{fig:flow-chart}. Because the
\texttt{smuggle()} function is defined in the notebook namespace, it
is also possible (though never necessary) to call it
directly. Deactivating \texttt{Davos} will delete the name
``\texttt{smuggle}'' from the namespace, unless its value has been
overwritten and no longer refers to the \texttt{smuggle()}
function. It will also deregister the \texttt{Davos} parser from the
set of input transformers run when each notebook cell is
executed.

\section{Illustrative Example}\label{sec:illustrative-example}

\begin{figure}[tp]
\centering
\includegraphics[width=\textwidth]{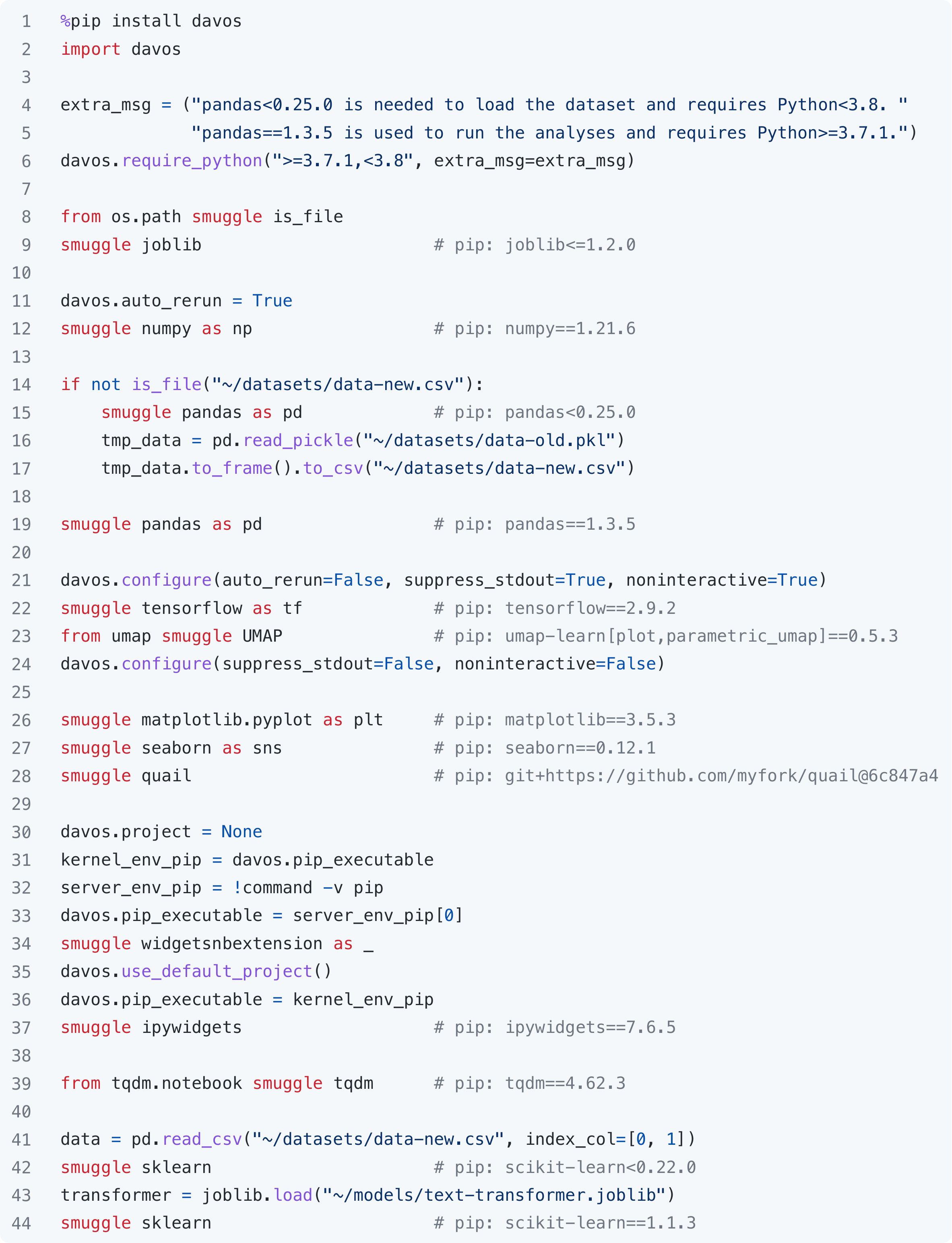}
\caption{\small \textbf{Example use case for \texttt{Davos}.}
  Snippets from this example are also excerpted in the main text of
  Section~\ref{sec:illustrative-example}.}
\label{fig:illustrative-example}
\end{figure}

The example code throughout Section~\ref{subsec:onion} illustrates a typical use case that we envision for \texttt{Davos}: a series of smuggle statements and onion comments with version specifiers or other options collectively describes and automatically constructs a reproducible environment for running the code that follows it.
When added to the top of a Jupyter notebook, this allows researchers to bundle their code and its dependencies into a single file that can be easily shared and run without any additional tools or setup, automatically installs its required packages at runtime, isolates them from the user's main Python environment, and ensures their versions do not change unexpectedly over time.
In this section, we have contrived a more complex scenario to highlight some of \texttt{Davos}'s more advanced features, and illustrate how they may be used to handle certain challenges that can arise when writing, running, and sharing reproducible scientific code.

Across different versions of a given package, various modules, functions,
and other objects may be updated, removed, renamed, or otherwise altered. In
addition to changing the behaviors of active computations, these changes can
render saved objects created using one version of a package incompatible with
other versions of the same package. For example, the popular
\texttt{pandas}~\cite{McKi10} library originally included the \texttt{Panel} data
structure for storing 3-dimensional arrays. In version 0.20.0, however, the
\texttt{Panel} class was deprecated, and in version 0.25.0, it was removed
entirely. Suppose a user had a dataset stored in a \texttt{Panel} object
(created using an older version of \texttt{pandas}) and had saved it to their
disk (e.g., for later reuse or to share with other users) by serializing the
\texttt{Panel} with Python's \texttt{pickle} protocol. The \texttt{pickle}
protocol is a popular built-in method of persisting data in Python that allows
users to save, share, and load arbitrary objects. However, in order to
successfully ``unpickle'' (i.e., load and restore) a ``pickled'' (i.e., previously saved)
object, that object's class must be defined in and importable from the same
module as it was when the object was originally saved. Thus, because of the \texttt{Panel} class's
removal, the user's dataset could not be read by any version of \texttt{pandas}
from 0.25.0 onward. These incompatibilities are also not limited solely to
traditional forms of data. For example, saved model states and other objects
may reference modules, functions, attributes, classes, or other objects that
may not be identical (or even present) across all versions of their associated
packages.

The example provided in Figure~\ref{fig:illustrative-example} demonstrates how
\texttt{Davos} can be used to circumvent these incompatibilities by
temporarily switching between different versions of the same package within a single runtime.
The example shows how a dataset and model that require
now-incompatible components of the \texttt{pandas} and
\texttt{scikit-learn}~\cite{PedrEtal11} libraries can be loaded in (using older
versions of each package) and used alongside more recent versions of each
package that provide new and improved functionality. When included at the top
of a Jupyter notebook, the code in Figure~\ref{fig:illustrative-example}
ensures that these objects will be loaded successfully and analyzed using the
same set of package versions no matter when or by whom the notebook is run.

After installing and importing \texttt{Davos} (lines 1--2), we first use the \texttt{davos.re\-qui\-re\_\-py\-thon()} function to constrain the Python version used to run the notebook (see Sec.~\ref{subsec:toplevel}).
As described above, the example code in Figure~\ref{fig:illustrative-example} loads two different versions of the \texttt{pandas} library: first, an older version needed to access a dataset saved in an outmoded format, then a newer one to use throughout the remainder of the notebook.
We therefore want to make sure upfront (in line 6) that the notebook's Python version falls within the range of versions that both of these two versions of \texttt{pandas} support.
If it does not, the function in line 6 will raise an error that includes a message to this effect (lines 4--5).
\begin{center}
\includegraphics[width=0.9\textwidth]{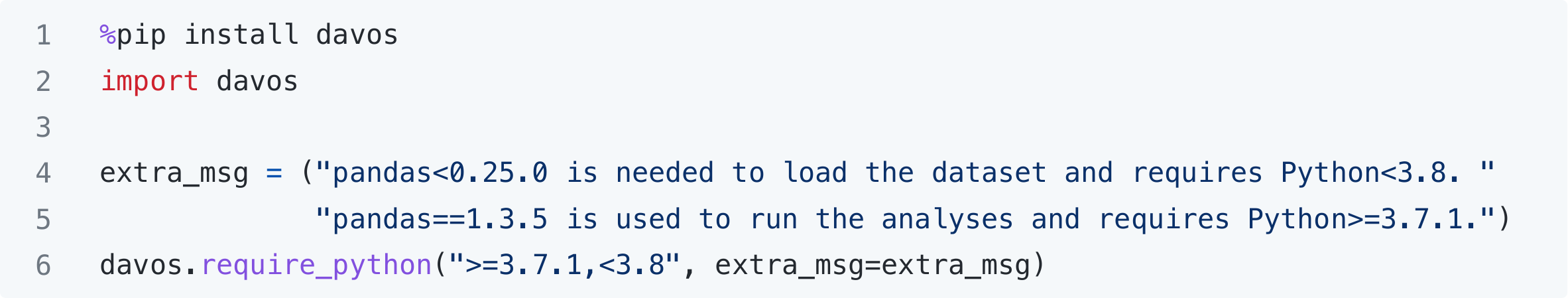}
\end{center}

Next, in lines 8--9, we \texttt{smuggle} two
utilities for interacting with local files in the code below. The
\texttt{smuggle} statement in line 8 loads the \texttt{is\_file()}
function from the Python standard library's \texttt{os.path}
module. Standard library modules are included with all Python
distributions, so this line is functionally equivalent to an
\texttt{import} statement and does not need or benefit from an onion
comment (since there is no chance the module will need to be installed).
Line 9 then loads the \texttt{joblib} package~\cite{Varo10},
installing it into the notebook's project directory if necessary. Since \texttt{joblib}'s I/O
interface has historically remained stable and backwards-compatible
across releases, requiring a particular exact version
would likely be unnecessarily restrictive. However, it is possible a
\textit{future} release could introduce some breaking change. The
onion comment in line 9 helps ensure that the analysis notebook will continue
to run properly in the future by limiting allowable versions to those
already released when the code was written:
\begin{center}
\includegraphics[width=0.9\textwidth]{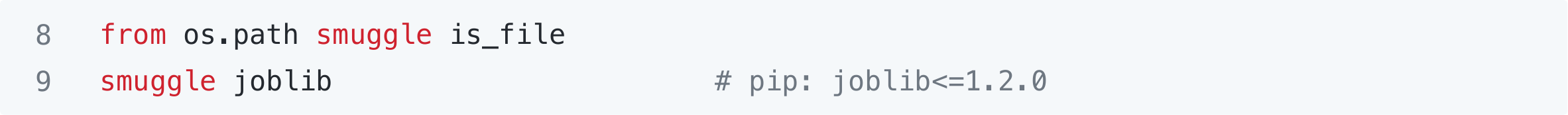}
\end{center}
It is worth noting, however, that beyond illustrative purposes, the benefit of specifying only a maximum version for \texttt{joblib} rather than an exact version is relatively minor.
The main advantage to relaxing a version constraint in an onion comment (when a package's behavior does not differ meaningfully between versions) is that doing so increases the likelihood that a satisfactory version will already be available in the user's Python environment, and therefore \texttt{Davos} will not need to install a new copy in the notebook's project directory.
For large packages, this can be a worthwhile consideration; however \texttt{joblib} is very lightweight---less than 0.5 MB pre-built, with no required dependencies.
Thus a more conservative approach that guarantees an exact version is used would also be reasonable in this case.

Line 11 then enables
\texttt{Davos}'s \texttt{auto\_rerun} option (see Sec.~\ref{subsec:config}) before smuggling the next
two packages: \texttt{NumPy} and
\texttt{pandas}. Because these packages rely heavily on custom C data
types, loading the particular versions specified in their onion comments may
require restarting the notebook kernel if different versions were previously
imported during the same interpreter session---including internally by other packages.
Enabling \texttt{auto\_rerun} allows \texttt{Davos} to handle kernel restarts automatically and continue running the code seamlessly without user intervention.
\begin{center}
\includegraphics[width=0.9\textwidth]{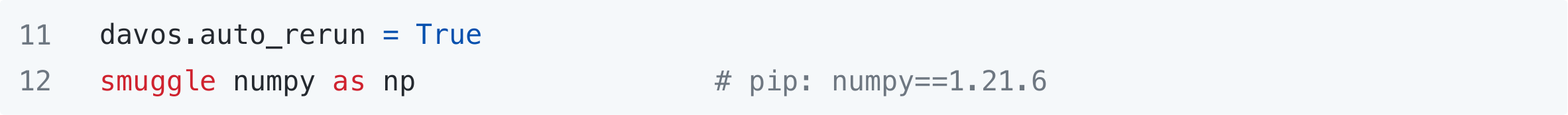}
\end{center}
In the case of \texttt{NumPy}, whether or not a kernel restart is necessary will depend on the user's existing Python environment.
The \texttt{joblib} package has an optional dependency on \texttt{NumPy} for memoizing and parallelizing array operations, and will \texttt{import numpy} internally to enable these features if the package is available.
If the user already has \texttt{NumPy} installed in their Python environment when \texttt{joblib} is smuggled in line 9, their installed version is different from the one specified in the onion comment on line 12, and there were changes made to \texttt{NumPy}'s C extensions between those two versions, then \texttt{Davos} will automatically restart the kernel and re-run the lines above.
The newly smuggled version would then be used both in the notebook itself and by \texttt{joblib} internally.

The primary reason for enabling the \texttt{auto\_rerun} option, however, is to manage the installation of \texttt{pandas} in the next lines:
\begin{center}
\includegraphics[width=0.9\textwidth]{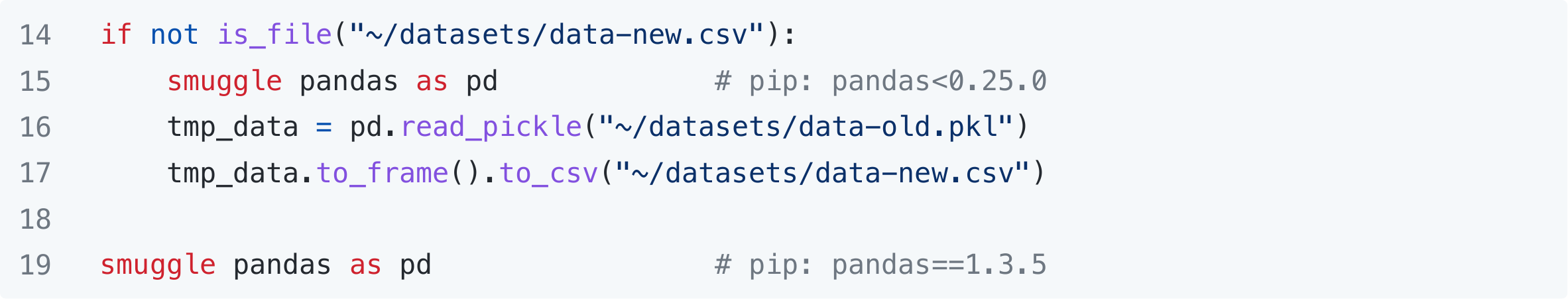}
\end{center}
If we suppose that the ``\texttt{data-old.pkl}'' file contains a dataset
stored in a pickled \texttt{Panel} object, then we must use a version of
\texttt{pandas} prior to v0.25.0 (i.e., the version in which the \texttt{Panel}
class was removed) to be able to read it. Line 15 ensures
that a sufficiently old version of \texttt{pandas} will be imported, enabling
the data to be successfully loaded in line 16 and (in line 17) written to a CSV
file, which can be read by any \texttt{pandas} version.

Newer versions of \texttt{pandas} have brought substantial improvements
including performance enhancements, bug fixes, and additional functionality. Although
the original dataset had to be read in using an older version of the package,
we can take advantage of these more recent updates by smuggling \texttt{pandas}
a second time in line 19 (whose onion comment specifies that version 1.3.5
should be installed and loaded). Since a different \texttt{pandas} version
has already been loaded by the Python interpreter (line 15) and there have been
substantial changes to the library (including its extension modules)
between that version and v1.3.5, the notebook
kernel must be restarted in order to fully unload the old version in favor of
the new one.
When \texttt{Davos} automatically does so and re-runs the code above, having now converted the dataset to a CSV file means the old version does not need to be reinstalled (line 14).

Next, line 21 uses the \texttt{davos.configure()} function to disable
the \texttt{auto\_rerun} option and simultaneously enable two other
options: \texttt{suppress\_stdout} and \texttt{noninteractive}. With
these options enabled, lines 22--23 \texttt{smuggle}
\texttt{TensorFlow}~\cite{AbadEtal15}, a powerful end-to-end platform
for building and working with machine learning models, and
\texttt{UMAP}~\cite{McInEtal18}, a package that implements a family
of related manifold learning techniques. The onion comment in line 23
also specifies that \texttt{UMAP} should be installed with the
optional requirements needed for its ``plot'' and ``parametric\_umap''
features. Together, these two packages depend on 36 other unique
packages, most of which have dependencies of their own. If many of
these are not already installed in the user's environment, lines
22--23 could take several minutes to run. Enabling the
\texttt{noninteractive} option ensures that the installation will
continue automatically without user input during that time. Enabling
\texttt{suppress\_stdout} also suppresses console outputs while installing
these packages and their many dependencies to prevent other potentially important outputs from being buried.
\begin{center}
\includegraphics[width=0.9\textwidth]{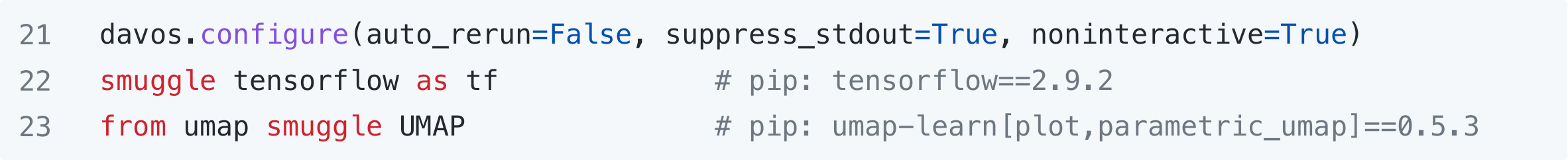}
\end{center}

After reverting these two options (line 24) to their default values, we next \texttt{smuggle}
specific versions of three plotting packages:
\texttt{Matplotlib}~\cite{Hunt07}, \texttt{seaborn}~\cite{Wask21}, and
\texttt{Quail}~\cite{HeusEtal17} (lines 26--28). Because the first two
are requirements of \texttt{UMAP}'s optional ``plot'' feature, they
will have already been installed (if necessary) by line 23, though possibly as
different versions than those specified in the onion comments on lines
26 and 28. If the installed and specified versions are the same, these
\texttt{smuggle} statements will function like standard \texttt{import}
statements to load the packages into the notebook's namespace. If they
differ, \texttt{Davos} will download the requested versions in place
of the installed versions, ensuring that they are used both in the notebook itself and by \texttt{UMAP} internally.
\begin{center}
\includegraphics[width=0.9\textwidth]{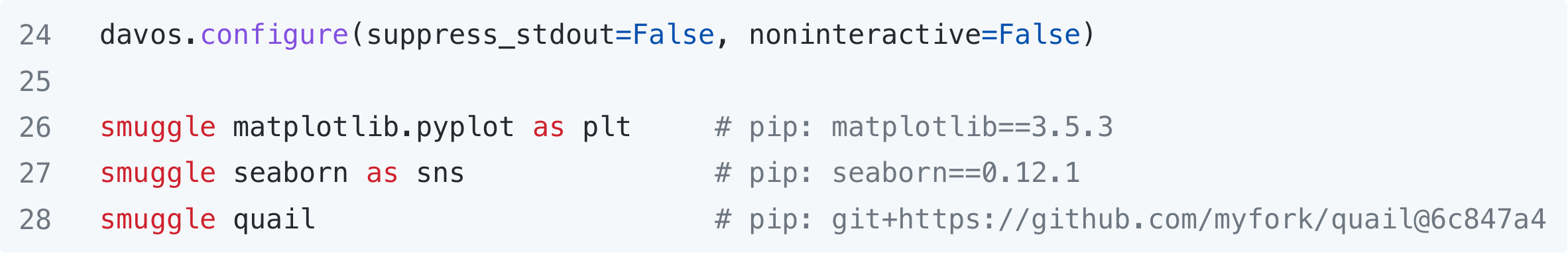}
\end{center}
The onion comment in line 28 specifies that \texttt{Quail} should be
installed from a fork of its GitHub repository (\texttt{myfork}), in its state as of a specific commit (\texttt{6c847a4}).
This ability to load packages directly from remote (or local) Git repositories can
enable developers to more easily use forked or customized versions of other
packages in their code, even if those versions have not been
officially released. Targeting specific VCS references (e.g., commits, tags, etc.) can also provide even finer-grained control over smuggled package versions than is possible with traditional version specifiers.

In lines 30--37, we demonstrate another aspect of \texttt{Davos}'s functionality that supports more advanced installation scenarios.
The \texttt{ipywidgets}~\cite{FredEtal15} package (also known as Jupyter Widgets) provides a Python API for creating interactive JavaScript widgets within a notebook.
It depends on the \texttt{widgetsnbextension} package, which provides the JavaScript machinery needed by the notebook frontend to display these widgets.
A complication is that \texttt{ipywidgets} must be installed in a location that is accessible from the IPython kernel (i.e., the Python runtime within the notebook itself), while \texttt{widgetsnbextension} must be installed in the environment that houses the Jupyter notebook server (a separate Python runtime that serves and manages the notebook frontend client).
In many basic setups, the IPython kernel and notebook server exist in the same environment.
However, a common ``advanced'' approach entails running the notebook server from a base environment, with additional environments each providing their own separate, interchangeable IPython kernels.

Lines 30--37 account for both of these possibilities programmatically:
\begin{center}
\includegraphics[width=0.9\textwidth]{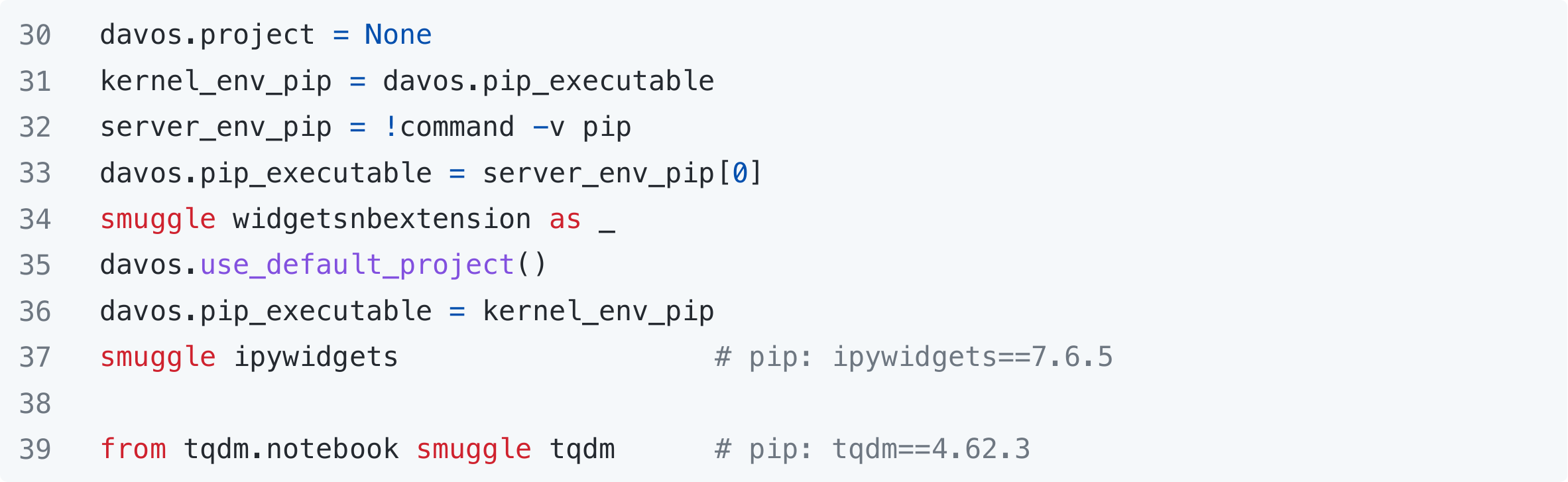}
\end{center}
First, in line 30, we set the \texttt{davos.project} attribute to \texttt{None} to temporarily
allow installing smuggled packages outside of the notebook's project directory.
As noted in Section~\ref{subsec:projects}, this is typically discouraged, as doing so can risk interfering with the user's Python environment if existing package versions are overwritten.
In this particular case, however, a combination of factors make this relatively safe and inconsequential.
First, the package we need to install directly into the notebook server environment (\texttt{widgetsnbextension}) is smuggled without an accompanying onion comment (line 34), meaning that \texttt{Davos} will not replace any version the user may already have installed.
Second, the package has no dependencies of its own, so if \texttt{Davos} does install it, no other packages will be installed or updated as a side effect.
Third, the package itself provides no functionality outside of rendering Jupyter widgets, so its presence would not alter any other code's expected behavior.

Next, in lines 31--33, we change the \texttt{pip} executable \texttt{Davos} uses to install smuggled packages (see Sec.~\ref{subsec:config}), storing the default executable's path in a variable before doing so.
When \texttt{Davos}'s project system is disabled, using a \texttt{pip} executable from a particular Python environment will cause smuggled packages to be installed into (and subsequently loaded from) that environment.
The default \texttt{pip\_executable} will install packages into the environment used to run the IPython kernel.
Here, the new value assigned to \texttt{davos.pip\_executable} in line 33 is the output of running ``\texttt{command -v pip}'' as a \mbox{\texttt{!}-prefixed} IPython system shell command in line 32 (``\texttt{command -v}'' outputs the path to an executable, similar to ``\texttt{which}'' but more portable).
IPython system shell command are always executed in the notebook server's environment---which may or may not be different from the kernel's environment---so this command's output will be the path to the server environment's \texttt{pip} executable.

After smuggling the \texttt{widgetsnbextension} package in line 34, we use the \texttt{davos.use\_\-default\_\-project()} function in line 35 to revert to installing package into the notebook's project directory, restore the default value of \texttt{davos.pip\_executable} in line 36, and \texttt{smuggle} the specified version of \texttt{ipywidgets} in line 37.
With these two packages now installed
and imported, line 39 smuggles \texttt{tqdm}~\cite{daCoEtal22}, which
displays progress bars to provide status updates for running code. In
Jupyter notebooks, the \texttt{tqdm.notebook} module can be imported
to enable more aesthetically pleasing progress bars that are displayed via
\texttt{ipywidgets}, if that package is installed and
importable. Therefore, to take advantage of this feature, it was
important to first ensure ensure that both \texttt{ipywidgets} and \texttt{widgetsnbextension} were available.

Next, we load in the reformatted dataset (line 41) and pre-trained
model (line 43) that we wish to use in our analysis. In our
hypothetical example, we can suppose that the model was provided as a
\texttt{scikit-learn} \texttt{Pipeline} object that passes data
through two pre-trained models in succession. First, a trained \texttt{CountVectorizer}
instance converts text data to an array of word counts. The
word counts are then passed to a topic model~\cite{BleiEtal03} using a
pre-trained \texttt{Latent\-Dir\-ich\-let\-Allocation} instance.
\begin{center}
\includegraphics[width=0.9\textwidth]{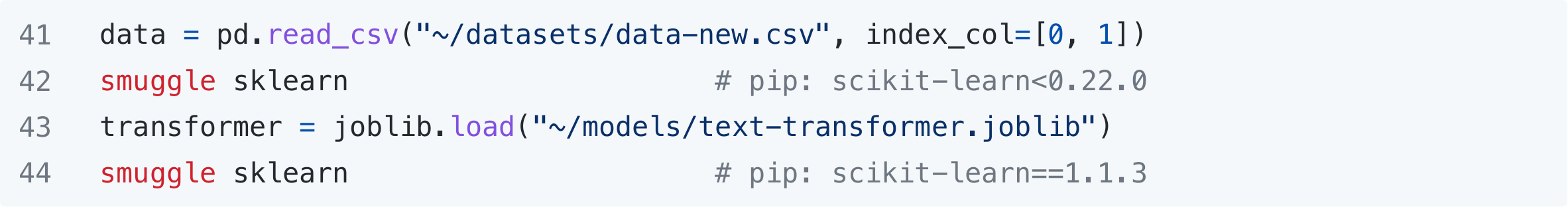}
\end{center}
Let us suppose that the \texttt{Pipeline} object had been saved by its
original creator using the \texttt{joblib} package, as
\texttt{scikit-learn}'s documentation recommends~\cite{skle22}. Because
\texttt{joblib} uses the \texttt{pickle} protocol internally, the
ability to save and load pre-trained models is not guaranteed across
different \texttt{scikit-learn} versions. For example, suppose that
the \texttt{Pipeline} object was created using \texttt{scikit-learn}
v0.21.3. In that version (and previous versions) of \texttt{scikit-learn}, the
\texttt{LatentDirichletAllocation} class was defined in
the \texttt{sklearn.de\-comp\-o\-si\-tion.online\_lda} module. However, in version
0.22.0, that module was renamed to ``\texttt{\_online\_lda}'' and in
version 0.22.1, it was again renamed to ``\texttt{\_lda}.''

In order to successfully load the model that includes the pre-trained
\texttt{Latent\-Dir\-ich\-let\-Allocation} instance, in line 42, we first
\texttt{smuggle} a version of \texttt{scikit-learn} prior to v0.22.0 (i.e.,
before the first time the relevant module's name was changed). Once
the model is loaded and reconstructed in memory from a compatible
package version (line 43), we upgrade to a newer version of
\texttt{scikit-learn} in line 44. Taken together, the code in
Figure~\ref{fig:illustrative-example} shows how \texttt{Davos} can
enable users to load in data and models that are incompatible with
newer versions of \texttt{pandas} and \texttt{scikit-learn}, but still
\textit{analyze} and manipulate the data and model output using the
latest approaches and implementations.

\section{Impact}\label{sec:impact}

We designed \texttt{Davos} for use in research settings, where code for numerous different tasks---from processing data, to running statistical analyses, to generating figures and tables for publication---is frequently shared between collaborators while working on a project, and eventually with the broader scientific community and general public upon its completion.
In these contexts, ensuring that shared code yields consistent, reproducible outputs across users and over time is critical, yet the tools available to researchers for doing so can be complex to set up and challenging to properly use.
This has the dual effect of discouraging scientists from sharing their code in a reproducible way (or at all), and making it significantly harder (or impossible) for others to successfully reproduce their results, adopt or extend their methods, and contribute to or build upon their work.
Ultimately, the need to install and master additional tools in order to share and run reproducible code can impede progress both on the individual level and on the broader scientific level.

While the \texttt{Davos} package by no means offers a universal solution to this problem, it can, in many cases, provide an effective yet more accessible alternative to existing systems for sharing reproducible Python code, such as virtual environments, containers, and virtual machines (Fig.~\ref{fig:code-sharing}).
For researchers, this can lower barriers to collaborating with peers and contributing to publicly available open science resources.
And by eliminating most of the setup costs of reconstructing the original researchers' computing environment(s), \texttt{Davos} also lowers barriers to entry for members of the scientific community and the public who seek to run shared code.

Among common systems for sharing reproducible Python code (see Fig.~\ref{fig:code-sharing}), \texttt{Davos} is most comparable to virtual environments in that it provides a lightweight mechanism for specifying and sharing a complete set of packages (and specific package versions) required by a particular project, and installing those packages into an isolated directory on the user's file system.
However, when used in conjunction with Jupyter notebooks, \texttt{Davos} offers a number of advantages over standard virtual environments that make it both easier to use and more effective at ensuring reproducibility of shared code.

First, \texttt{smuggle} statements and onion comments enable researchers to specify their notebooks' dependencies directly within the code that requires them (see Secs.~\ref{subsec:smuggle} and~\ref{subsec:onion}).
This eliminates the need to either manually create and maintain separate configuration files or use an additional tool to generate them, and to then distribute these files alongside their primary code base for users to download.

Second, \texttt{Davos} automatically checks for, installs, isolates, and imports any required packages at runtime.
This allows users to download and immediately run a \texttt{Davos}-enhanced notebook without any prerequisite setup (beyond that needed to run a standard Jupyter notebook).
By contrast, before running a notebook whose dependencies are managed via a virtual environment, the user would first need to run a series of shell commands to manually create a new environment, populate it with packages from the researcher's configuration file (which the user must also have obtained), and then either use the \texttt{ipykernel} package to register the environment as a Jupyter kernel, or activate and deactivate the environment before and after (respectively) each time the notebook was launched.
Beyond reducing this initial setup cost, \texttt{Davos}'s runtime-based approach to dependency management affords a second important benefit.
While creating and configuring a virtual environment ensures that a specific set of packages is initially installed, it does not guarantee that they will \textit{remain} installed after that point.
For example, a researcher who creates a virtual environment in which to run a set of data analyses---or a different user who later recreates that environment to reproduce them---might at some point install an additional package into the environment after its initial setup (e.g., to implement a new analysis idea).
Depending on the requirements of this new package, this could cause one or more initially installed packages to be upgraded or downgraded to a different version.
If the individual does not happen to notice this change when it occurs, differences between those packages' expected and installed versions may introduce bugs into previously written code or subtly alter its output when it is next run.
Because \texttt{smuggle} statements and onion comments are evaluated every time a \texttt{Davos}-enhanced notebook is run, they function to ensure that the notebook's dependencies are always satisfied and that any such inadvertent changes would be automatically caught and corrected.

Third, running a shared notebook that uses \texttt{Davos} to manage its dependencies often requires less (but never more) space on the user's system than running an identical notebook inside a virtual environment.
While typical virtual environment directories will contain \textit{all} requirements listed in their configuration files, \texttt{Davos}'s isolated project directories (by default; see Sec.~\ref{subsec:projects}) contain only those not already available in the user's existing Python environment.
This is another feature made possible by \texttt{Davos}'s runtime-based dependency management system.
\texttt{Davos} can safely draw from packages in the user's main environment to satisfy a notebook's dependencies, when possible, because if those packages were to be updated or removed at any point such that they no longer met the notebook's requirements, appropriate versions would be installed into the project directory the next time the notebook was executed.

Finally, neither virtual environments nor \texttt{Davos}-enhanced notebooks can bundle specific versions of Python with which to run shared code.
However, \texttt{Davos}'s \texttt{require\_\-python()} function provides a simple mechanism for indicating a required or constrained Python version and alerting the user at runtime (by raising an exception) if their Python version is incompatible.
In terms of \texttt{Davos}'s ability to ensure that shared code can be executed reproducibly by other users, this falls short of the capabilities of more complex tools that can provide complete copies of Python (Fig.~\ref{fig:code-sharing}).
However, the functionality \texttt{Davos} provides over what is possible with a standard virtual environment is to remove the \textit{expectation} that running a particular notebook will reproduce an expected outcome in situations where this is either impossible or not guaranteed.
With this understanding, a user may choose to install a compatible Python version through some other means or elect to still run the code, but will not be surprised by a potential failure to execute successfully or output an expected result.

Beyond research applications, \texttt{Davos} is also useful in
pedagogical settings. For example, in programming courses, instructors
and students may use the \texttt{Davos} package to ensure their
notebooks will run correctly on others' machines. When combined with
online notebook-based platforms like Google Colaboratory,
\texttt{Davos} provides a convenient way to manage dependencies within
a notebook without requiring any software (beyond a web browser) to
be installed on the students' or instructors' systems. For the same
reasons, \texttt{Davos} also provides an elegant means of sharing
ready-to-run notebook-based demonstrations or tutorials that install
their dependencies automatically.

Since its initial release, \texttt{Davos} has found use in a variety of applications. In
addition to managing computing environments for multiple prior and ongoing
research studies~\citep{MannEtal23a, OwenMann23, ZimaEtal23}, \texttt{Davos} is being
used by both students and instructors in programming and research methods courses such
as Storytelling with Data~\cite{Mann21a} (an open course on data science,
visualization, and communication), Laboratory in Psychological
Science~\cite{Mann22} (an open course on experimental and statistical methods
for psychology research), and the Methods in Neuroscience at Dartmouth (MIND)
Computational Summer School~\cite{MIND23} (a week-long intensive course on
computational neuroscience methods) to simplify distributing lessons
and submitting assignments, as well as in online demos such as
\texttt{abstract2paper}~\cite{Mann21b} (an example application of
GPT-Neo~\cite{GaoEtal20, BlacEtal21}) to share ready-to-run code that installs
dependencies automatically. The 2023 offering of Neuromatch
Academy~\cite{vanVEtal21} also included an ``experimental'' module that uses
\texttt{Davos} to manage dependencies related to a large language model-based
tutor~\cite{MannEtal23b}.

Our work also has several more subtle ``advanced'' use cases and potential
impacts. Whereas Python's built-in \texttt{import} statement is agnostic to
packages' version information, \texttt{smuggle} statements (when combined with
onion comments) are version-sensitive. And because onion comments are parsed at
runtime, required packages and their specified versions are installed in a
just-in-time manner. Thus, it is possible in most cases to \texttt{smuggle} a
specific package version or revision even if a different version has already
been loaded. This enables more complex uses that take advantage of multiple
versions of a package within a single interpreter session (e.g., see
Sec.~\ref{sec:illustrative-example} and Fig.~\ref{fig:illustrative-example}).
This could be useful in cases where specific features are added or removed from
a package across different versions, or in comparing the performance or
functionality of particular features across different versions of the same
package.

A second more subtle impact of our work is in providing a
proof-of-concept of how the ability to add new ``keyword-like''
operators to the Python language could be specifically useful to
researchers. With \texttt{Davos}, we accomplish this by leveraging
IPython notebooks' internal code parsing and execution machinery. We
note that, while other popular packages similarly use these mechanisms
to providing notebook-specific functionality (e.g.,
\cite{Hunt07,HeusEtal18}), this approach also has the potential to be
exploited for more nefarious purposes. For example, a malicious user
could design a Python package that, when imported, substantially
changes the notebook's functionality by adding new \textit{unexpected}
keyword-like objects (e.g., based around common typos). We also note
that this implementation approach means \texttt{Davos}'s functionality
is currently restricted to IPython notebook environments. However,
there have been early-stage discussions of providing this sort of
syntactic customizability as a core feature of the Python language itself,
including a draft proposal~\cite{Shan20}. In addition to enabling
\texttt{Davos} to be extended for use outside of notebooks, this could
lead to exciting new tools that, like \texttt{Davos}, extend the
Python language in useful and more secure ways.

\subsection{Pitfalls and limitations}

While \texttt{Davos} enables developers to conveniently specify all project
dependencies, there are some edge cases and limitations that are worth
considering.
First, prior studies on reproducibility of Jupyter notebooks~\cite[e.g.,][]{PimeEtal19} identified a key challenge in the fact that, unlike Python scripts, notebook cells may be manually executed in an arbitrary order, and therefore potentially in a different order than they were executed by the notebook's original author.
This can result in situations where, for example, a cell's execution fails because its code calls a function that has not yet been defined, or accesses a variable that refers to a different object than is expected at that point in the notebook.
In theory, using \texttt{Davos} to \texttt{smuggle} multiple versions of the same package in different cells of a notebook could exacerbate this issue if a user executed those cells out of their intended order, such that their currently imported version of a core dependency was different from what a particular cell expected or required.
Therefore, an important consideration when using \texttt{Davos} to facilitate complex, multi-package-version runtimes in this way is that executing notebook cells in order is perhaps even more important than it would be in a standard (i.e., non-\texttt{Davos}-enhanced) notebook.
While (as noted in Sections~\ref{sec:illustrative-example} and~\ref{sec:impact}) we consider this an ``advanced feature'' of Davos rather than typical usage, we propose a relatively simple set of ``best practices'' that substantially mitigate the risk of creating ambiguous states within a notebook.
First, any \texttt{Davos}-enhanced notebook (or simply any notebook) that is intended to be run by more than one individual should be organized with its code cells in their intended execution order from top to bottom.
If an edge case arises in which this is not possible, the intended order should be clearly indicated in code comments and/or markdown cells.
Second, when smuggling multiple different versions of a package within a notebook, one version of the package may be designated the ``main'' version, and any others designated as ``alternate'' versions.
The main version should be the primary version used throughout the notebook, while alternates are those temporarily required for a specific task or functionality.
For example, in Figure~\ref{fig:illustrative-example}, \texttt{pandas} v1.3.5 and \texttt{scikit-learn} v1.1.3 are the main versions of their respective packages as they are used throughout the remainder of the code once they are loaded.
Meanwhile \texttt{pandas<0.25.0} and \texttt{scikit-learn<0.22.0} are alternate versions because they are temporarily smuggled for the specific purpose of loading an outmoded dataset and model and then immediately replaced with main versions after their use is complete.
Any time an alternate package version is needed, the \texttt{smuggle} statement used to install and load it, the operations it is required to perform, and a second \texttt{smuggle} to (re-)install and load the main package version should all be contained within a single notebook cell.
This ensures that (barring other unrelated errors in the cell's execution) the main version will always be installed and imported when any given notebook cell is run.
In other words, in Figure~\ref{fig:illustrative-example}, lines 14--19 should be run within a single cell, and lines 42--44 should also be run in a single cell.

A second limitation of \texttt{Davos} relates to how packages are installed and managed.
As of this writing, \texttt{Davos} can install packages using \texttt{pip}, but not
other standard Python package management systems such as \texttt{conda}~\cite{Anac12}.
Therefore packages that are not installable via \texttt{pip} are currently
unsupported by \texttt{Davos}. We anticipate adding support for other package management
systems, including \texttt{conda}, in a future release.  Because \texttt{Davos} relies on
\texttt{pip} to install packages, it is also subject to the same limitations as
\texttt{pip} itself. For example, \texttt{pip}-installing a package that depends on
a previously \texttt{smuggle}d package may result in the previously smuggled package
being upgraded or downgraded to a different version. Whereas lockfiles, or lockfile-based
systems like \texttt{Poetry}~\cite{Eust19}, place stronger guarantees that each package will
have a stable version, we have opted for a more flexible (but, consequently, less
deterministic) implementation for \texttt{Davos}.  This enables us to support
more advanced use cases, such as those described in Section~\ref{sec:illustrative-example},
but at the cost of managing potential conflicts between \texttt{smuggle}d
packages.

A third limitation of \texttt{Davos} is that it cannot be used to manage projects that
depend on non-Python software. For example, system software or libraries from
other languages (e.g., in a mixed Python and R notebook), cannot be
\texttt{smuggle}d by \texttt{Davos}. A notebook that utilizes or depends on non-Python
software would therefore need to use existing non-\texttt{Davos} approaches to managing
those requirements.

\texttt{Davos}'s ``projects'' system (Sec.~\ref{subsec:projects}) provides a safe way
of managing project dependencies without interfering with the user's Python
environment. By default, each \texttt{Davos}-enhanced notebook creates and uses its own
notebook-specific project directory, which is named based on the notebook's absolute
path. However, programmatically determining the path to the currently running
notebook may not be possible in some environments.
For example, \texttt{Davos} queries the Jupyter Server API to determine the notebook's name, but some non-browser applications may implement mechanisms for communicating with the IPython kernel without starting a Jupyter server.
As of this writing, \texttt{Davos} fully supports most common notebook environments, including classic Jupyter Notebooks,
JupyterLab, Google Colaboratory, Binder, Kaggle Notebooks, JetBrains IDEs (PyCharm, DataSpell, etc.), and Visual Studio Code, among others.
However, in cases where \texttt{Davos} fails to determine the current notebook's path, it will issue a warning and fall back to using a generic project named ``davos-fallback.'' This project is shared across all such occurrences and exists to ensure that even if some component of \texttt{Davos}'s project system fails, smuggling packages will still not affect the user's main Python environment. If this occurs, the user can also manually set \texttt{Davos} to use the ``normal'' default project for the current by setting \texttt{davos.project} to its absolute or relative path.

\section{Conclusions}

The \texttt{Davos} package supports reproducible research by providing
a novel, lightweight system for sharing notebook-based code. It
expands on Python's ``batteries included'' philosophy to enable running shared
notebooks with no setup required, and defines a simple, self-documenting format for
specifying dependencies in keeping with the ``literate programming'' paradigm that Jupyter notebooks support.
We designed \texttt{Davos} to fill a niche we believe will help facilitate contributing to and engaging with open science resources.
But perhaps the most exciting uses of the \texttt{Davos} package are those
that we have \textit{not} yet considered or imagined. We hope that the
research and scientific Python communities will find \texttt{Davos} to provide a convenient
means of managing project dependencies to facilitate code sharing and collaboration. We
also hope that some of the more advanced applications of our package
might lead to new insights or discoveries.

\section*{Author Contributions}

\textbf{Paxton C. Fitzpatrick}: Conceptualization, Methodology,
Software, Validation, Writing - Original Draft,
Visualization. \textbf{Jeremy R. Manning}: Conceptualization,
Resources, Validation, Writing - Review \& Editing, Visualization, Supervision,
Funding acquisition.

\section*{Funding}

Our work was supported in part by NSF grant number 2145172 to JRM\@.
The content is solely the responsibility of the authors and does not
necessarily represent the official views of our supporting
organizations.

\section*{Declaration of Competing Interest}

We wish to confirm that there are no known conflicts of interest
associated with this publication and there has been no significant
financial support for this work that could have influenced its
outcome.

\section*{Acknowledgements}

We acknowledge useful feedback and discussion from the students of
JRM's \textit{Storytelling with Data} course (Winter, 2022 offering)
who used preliminary versions of our package in several assignments,
and the students of the Methods in Neuroscience at Dartmouth (MIND)
Computational Summer School (2023 offering) who used our package
during several workshops and tutorials.

\bibliographystyle{elsarticle-num}
\bibliography{main}

\end{document}